\documentclass{article}

\usepackage[nonatbib, final]{neurips_2023}
\usepackage[numbers]{natbib}
\usepackage[utf8]{inputenc} 
\usepackage[T1]{fontenc}    
\usepackage[hidelinks]{hyperref}       
\usepackage{url}            
\usepackage{booktabs}       
\usepackage{amsfonts}       
\usepackage{nicefrac}       
\usepackage{microtype}      
\usepackage{xcolor}         
\usepackage{graphicx}
\usepackage{multirow}
\usepackage[normalem]{ulem}
\useunder{\uline}{\ul}{}
\usepackage[title]{appendix}
\usepackage[para]{footmisc}
\title{Unsupervised Segmentation of Colonoscopy Images}

\author{%
Heming Yao$^{1}
\thanks{equal contribution}
\protect\phantom{\footnotesize 1}\textsuperscript{,}
\thanks{corresponding authors: \{yao.heming, richmond.david\}@gene.com \quad}$
\quad Jérôme Lüscher$^{1}\footnotemark[1]$
\quad Benjamin Gutierrez Becker$^2$
\quad Josep Arús-Pous$^2$  \\
\textbf{Tommaso Biancalani}$^1$ \quad \textbf{Amelie Bigorgne}$^3$ \quad \textbf{David Richmond}$^{1}\footnotemark[2]$ \\
$^1$Biology Research | AI Development (BRAID), gCS, Genentech\\ $^2$Pharma Research \& Early Development, Data and Analytics, Roche\\ 
$^3$Technology and Translational Research Team, PDC, ION, Roche
}

\begin{document}

\maketitle

\begin{abstract}
Colonoscopy plays a crucial role in the diagnosis and prognosis of various gastrointestinal diseases. 
Due to the challenges of collecting large-scale high-quality ground truth annotations for colonoscopy images, and more generally medical images, we explore using self-supervised features from vision transformers in three challenging tasks for colonoscopy images.
Our results indicate that image-level features learned from DINO models achieve image classification performance comparable to fully supervised models, and patch-level features contain rich semantic information for object detection.
Furthermore, we demonstrate that self-supervised features combined with unsupervised segmentation can be used to `discover' multiple clinically relevant structures in a fully unsupervised manner, demonstrating the tremendous potential of applying these methods in medical image analysis.

\end{abstract}

\section{Introduction}
\label{Introduction}

Colonoscopy is the standard of care procedure for diagnosing many gastrointestinal diseases.
It is routinely used to identify mucosal features of inflammatory bowel disease (IBD), such as the loss of normal vascular pattern, erythema (reddening), ulcers and bleeding.
As a result, numerous machine learning methods have been developed for unbiased scoring of IBD severity \cite{gutierrez2021training, yao2021fully} and segmentation of abnormal structures, such as polyps in colon cancer \cite{xu2022deep,raju2023advanced,wang2023foundation}.
However, two key challenges have limited progress towards the goal of complete semantic segmentation of colonoscopy videos: 
(1) 
mucosal features are extremely challenging to define in a systematic way, as evidenced by low inter-rater agreement between gastroenterologists when evaluating mucosal inflammation \cite{de2004inter}, and
(2) dense annotation of colonoscopy videos is incredibly cumbersome and difficult to scale.
Furthermore, there is growing interest in the unsupervised discovery of new and unbiased mucosal features that go beyond existing definitions.
Thus, unsupervised approaches hold tremendous potential to impact the diagnosis, treatment and development of new therapeutics for colonic diseases.

Recently, unsupervised segmentation methods have made dramatic progress, driven by advances in self-supervised learning (SSL) combined with vision transformers (ViTs) \cite{melas2022deep,hamilton2022unsupervised,zhang2023dive} (see Appendix~\ref{sec:related_work} for Related Work).
In this study, we explore a practical application of these methods to the task of unbiased discovery of clinically relevant structures within colonoscopy images.
We systematically compare representations from multiple SSL methods for the task of image classification and object detection, and show promising results on a public benchmark colonoscopy dataset. 
We then apply an unsupervised segmentation workflow based on these SSL models, and show unsupervised discovery of 6 interpretable and clinically relevant mucosal features from an internal clinical trial dataset.
These findings demonstrate the tremendous potential of SSL and unsupervised segmentation applied to medical images, and could have numerous applications, including the automated quantification of treatment effect (i.e., characterization of tissues before vs after treatment), and stratification of patient populations for personalized healthcare.


\begin{figure}
\centering
\includegraphics[width=13cm]{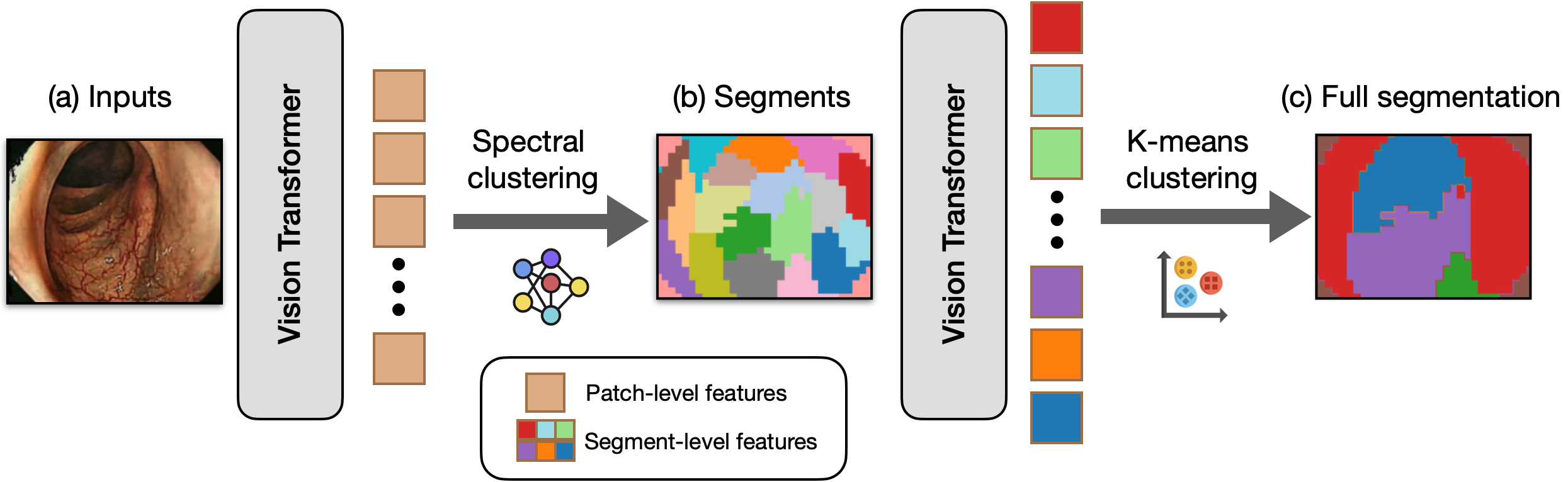}
\caption{Pipeline for unsupervised semantic segmentation (best viewed in color).}
\label{fig:method_step_images}
\end{figure}

\section{Methods}
\label{Methods}

We adopt the method proposed by Deep Spectral \cite{melas2022deep} for unsupervised segmentation using self-supervised ViTs.
The segmentation workflow has two stages, as illustrated in Figure \ref{fig:method_step_images}. 
In the first stage, we extract self-supervised patch-level features using pre-trained ViTs as well as low-level color features. 
We then compute the pairwise affinity between all patches, and apply spectral clustering on the affinity matrix to break each image into semantic segments (Figure \ref{fig:method_step_images}b).
In the second stage, we feed each segment back into the ViT to generate segment-level features, which are then clustered across images using K-means clustering to generate semantically consistent segmentations across the dataset.
Finally, we manually review the clusters to identify interpretable concepts associated with each cluster. The overall workflow is fully unsupervised, aside from the final review.

The workflow described above is based on the insight that ViTs learn rich patch-level semantic features, even in networks trained with image-level contrastive objectives \cite{amir2021deep}.
We conduct a comparative analysis of
ViTs trained with a combination of different frameworks and datasets to identify which approach yields features that are most transferrable to unsupervised medical segmentation.
We compare the following: 
(i) ViTs trained on our internal dataset of over 500k colonoscopy images (referred to as "Etro") using the DINO v1 framework \cite{caron2021emerging} 
(ii) publicly released models trained on ImageNet using DINO v1, DINO v2 \cite{oquab2023dinov2} and MAE \cite{MaskedAutoencoders2021}, and
(iii) a recently released endoscopy foundation model (Endo-FM) trained on 5M endoscopic images \cite{wang2023foundation}.
We refer to the models trained on endoscopy images (Etro, Endo-FM) as domain-specific models.
Finally, we evaluate the above pipeline on a series of increasingly challenging tasks described in the next section.





\section{Experiments}

In this section, we present experiments to evaluate the quality of self-supervised representations across 3 downstream tasks: (i) image classification, (ii) object detection, and (iii) unsupervised discovery of mucosal features.
We refer readers to Appendix \ref{sec:dataset} and \ref{sec:implementation} for details about the datasets and implementation, respectively.
Additional results are presented in Appendix \ref{sec:add_results}.

\label{Results}

\subsection{Image Classification}

We first evaluate the quality of our pretrained features on image classification tasks from the HyperKvasir benchmark dataset, using the evaluation protocol proposed by authors of DINO \cite{caron2021emerging}. 
Specifically, we report the performance from both linear probing and weighted K-nearest neighbor (KNN) classifiers on a 23-class classification task defined by the authors of HyperKvasir \cite{borgli2020hyperkvasir}.
We also evaluate performance of a separate 3-class classification task on a subset of images with Mayo Clinic Endoscopic Subscore (MCES) > 0.
MCES is the standard scale for evaluating the severity of inflammation in ulcerative colitis, and is a challenging task with high inter-rater variability \cite{yao2021fully}.


We report the performance of our best performing internal model (ViT-B/16 Etro), as well as results from Endo-FM \cite{wang2023foundation} and numerous models trained on ImageNet \cite{caron2021emerging}.
Overall, domain specific models perform best on the unsupervised MCES classification subtask that is most relevant for segmenting IBD-related mucosal features in an unsupervised manner (micro-F1: 67.5\%; Table \ref{tab:KNN}).
However, models trained on ImageNet surprisingly outperform domain specific models after linear probing is applied (Table \ref{image-classification-table}).
DINOv1 ViT-B/8 trained on ImageNet is the best performing model when evaluated with linear probing 
(micro-F1: 75.0\% and 89.8\% on 3-class and 23-class classification, respectively),
and MAE is consistently a little worse than DINO models trained on ImageNet.
Importantly, the best performing SSL models with linear probing are typically within 1-2\% of the published state-of-the-art performance achieved with fully supervised training, confirming that SSL features are indeed able to capture good representations for downstream colonoscopy video analysis.
We also note that our internal Etro model performs very similarly to Endo-FM, despite the fact that Endo-FM includes images from HyperKvasir in its training set.



\renewrobustcmd{\bfseries}{\fontseries{b}\selectfont}
\renewrobustcmd{\boldmath}{}
\newrobustcmd{\B}{\bfseries}

\begin{table}[]
\centering
\caption{\textbf{Image Classification Performance on HyperKvasir}. We report average macro-F1 (\%) and micro-F1 (\%) scores from two-fold cross-validation. Best SSL results are in bold. Fully supervised results are included for reference.}
  \label{image-classification-table}
\begin{tabular}{llcccc}
\hline
\multirow{2}{*}{Type of method}                                                            & \multirow{2}{*}{Model name} & \multicolumn{2}{c}{Full 23-class} & \multicolumn{2}{c}{MCES 3-class} \\
                                                                                           &                             & F1 (ma.)        & F1 (mi.)        & F1 (ma.)        & F1 (mi.)       \\ \hline
\multirow{2}{*}{\begin{tabular}[c]{@{}l@{}}Domain SSL \\ w/ linear probing\end{tabular}}   & DINOv1 ViT-B/16 (Etro)      & 57.4            & 87.5            & 70.6            & 73.9           \\
                                                                                           & DINOv1 ViT-B/16 (Endo-FM)   & 58.0            & 88.3            & 69.9            & 72.3           \\ \hline
\multirow{4}{*}{\begin{tabular}[c]{@{}l@{}}ImageNet SSL \\ w/ linear probing\end{tabular}} & DINOv1 ViT-B/8              & 59.4            & \textbf{89.8}      & \textbf{71.5}      & \textbf{75.0}     \\
                                                                                           & DINOv1 ViT-B/16             & 59.6            & 88.8            & 70.2            & 73.9           \\
                                                                                           & DINOv2 ViT-L/14             & \textbf{59.7}      & 89.6            & 70.4            & 74.0           \\
                                                                                           & MAE ViT-L/16                & 57.8            & 87.8            & 69.8            & 73.1           \\ \hline
Fully supervised                                                                           & DenseNet                    & 61.9   & 90.7   & 72.9   & 75.1  \\ \hline
\end{tabular}
\end{table}

\subsection{Object Detection}

Next, we evaluate the potential of self-supervised patch-level features for detecting annotated polyps from the HyperKvasir dataset, as a proxy task for detecting clinically relevant structures.
We first evaluate linear probing on segment-level features returned by our unsupervised segmentation pipeline and find that models trained with DINO on ImageNet achieve the best performance, with an average F1@IoU=0.3 exceeding 80\% and an average IoU exceeding 50\% (Table \ref{polyps}). 

We also evaluate unsupervised polyp segmentation, and observe that for some models pretrained with DINO on ImageNet, 
the F1 score and IoU is just slightly ($\sim$ 5\%) lower than obtained through linear probing. 
However, we also observe that the unsupervised methods are more prone to over-segmentation or under-segmentation of the polyps, and their performance is more sensitive to the choice of hyperparameters.
As a baseline comparison, we implement weakly supervised polyp segmentation using attention maps from the fully supervised DenseNet model in Table \ref{image-classification-table}.
The F1 score is around 75\%, lower than the performance of the best SSL models. 
The success of unsupervised polyp segmentation again highlights the potential of these methods for unsupervised semantic segmentation and mucosal feature discovery.
Examples of polyp segmentation results are presented in Figure \ref{fig:polyps}.


\begin{table}[bp]
\caption{\textbf{Polyp Detection and Segmentation Performance on HyperKvasir.} We report average F1 score @ IoU=0.3 (detection) and average IoU (segmentation) from two-fold cross-validation.}
\label{polyps}
\begin{tabular}{llcccc}
\hline
\multirow{2}{*}{Type of method} & \multirow{2}{*}{Model name} & \multicolumn{2}{c}{Linear Probing (\%)} & \multicolumn{2}{c}{Unsupervised (\%)} \\
                                &                             & F1                 & IoU                & F1                & IoU               \\ \hline
\multirow{2}{*}{Domain SSL}     & DINOv1 ViT-B/16 (Etro)      & 78.8               & 48.1               & 68.4              & 39.4              \\
                                & DINOv1 ViT-B/16 (Endo-FM)   & 72.8               & 41.4               & 50.7              & 28.3              \\ \hline
ImageNet SSL                    & DINOv1 ViT-B/8              & 84.2               & \textbf{56.0}      & 79.1              & \textbf{52.0}     \\
                                & DINOv1 ViT-B/16             & 81.6               & 51.9               & 76.9              & 47.3              \\
                                & DINOv2 ViT-L/14             & \textbf{84.7}      & 55.9               & \textbf{80.6}     & 50.1              \\
                                & MAE ViT-L/16                & 71.2               & 41.5               & 40.4              & 26.1              \\ \hline
\end{tabular}
\end{table}

\begin{figure}
\centering
\includegraphics[width=12cm]{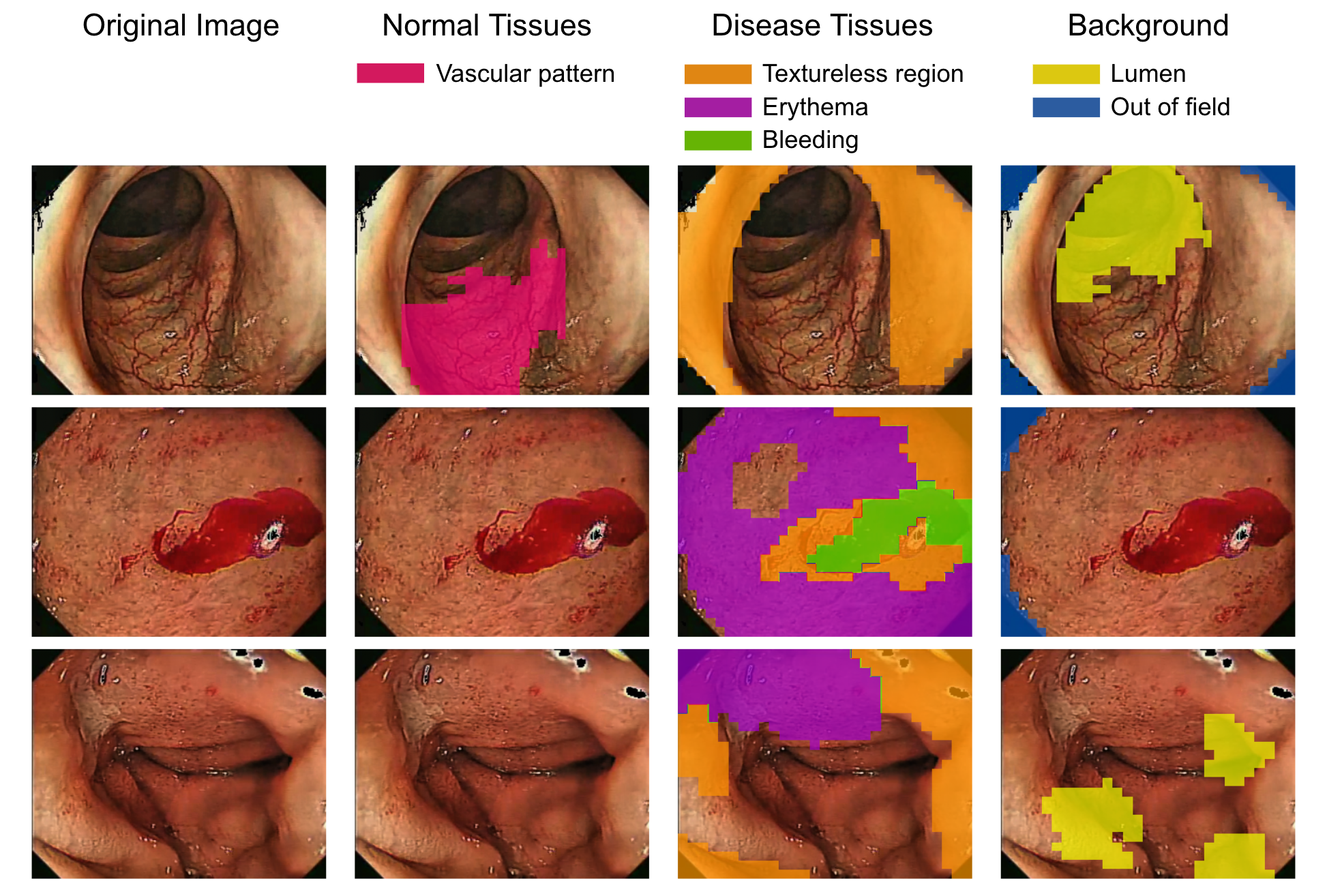}
\caption{\textbf{Unsupervised Discovery of Mucosal Features.} We show images overlaid with "discovered" concepts related to normal tissues, disease tissues, and background, respectively.}
\label{fig:concept}
\end{figure}

\subsection{Discovery of Mucosal Features}
Finally, we evaluate unsupervised semantic segmentation of the entire colonoscopy image, which is a much more challenging task compared to identifying salient objects, such as polyps.
We present proof-of-concept results on 20 images with varying levels of disease severity from our internal Etro  dataset.
We cluster semantic segments into 15 clusters using K-means, and manually identify six clusters that we are able to assign to interpretable concepts.
The clusters included two concepts associated with background (lumen, and out-of-field regions), one concept associated with normal tissues (normal vascular pattern), and three concepts indicative of disease severity (textureless regions, erythema, and bleeding). Figures \ref{fig:concept} and \ref{fig:semantic_seg_more} presents examples in which masks built from these concepts are overlaid onto the original images. The predicted masks are promising though not perfect.

We qualitatively observe that domain-specific models yield better results for unsupervised semantic segmentation of the mucosa. This aligns with our observation that the domain-specific models achieve better performance in MCES classification using a KNN classifier (Table \ref{tab:KNN}), despite performing worse at salient object detection.
Results shown in Figures \ref{fig:concept} and \ref{fig:semantic_seg_more} were generated using our internal Etro model.

\section{Conclusion}
In this study, we evaluated the use of SSL combined with ViTs across three common tasks in colonoscopy video analysis, with the ultimate goal of fully unsupervised semantic segmentation of colonoscopy images.
We observed promising results across all tasks. 
Surprisingly, for a general 23-class classification task, as well as salient object detection, ViTs pretrained with DINO on ImageNet performed the best, despite having never seen colonoscopy images during training.
For KNN classification of the MCES score, as well as unsupervised semantic segmentation, ViTs pretrained with DINO on large clinical datasets of colonoscopy videos performed best, suggesting that they are able to capture more clinically relevant mucosal features. We posit that the greater diversity of images that ImageNet models were trained on may enable them to extract generalizable features and perform well when applied to unseen datasets, whereas fine-tuning the model on domain-specific datasets may lead to catastrophic forgetting of some features.

To the best of our knowledge, this is the first work to demonstrate that mucosal tissues can be segmented in a fully unsupervised manner, and encourages further work to extract unbiased insights from colonoscopy videos, and more generally medical images.
Furthermore, the competitive results of off-the-shelf ViT models across all tasks also open the door to researchers without access to a large proprietary dataset of colonoscopy videos.
We refer readers to Appendix \ref{sec:limitation} for an extended discussion of limitations and future work.




\section*{Acknowledgments}
We would like to express our gratitude to Marco Prunotto, Stefan Fraessle, Daniela Bojic, Emily Fisher, KT Park, Mahtab Bigverdi, Burkhard Hoeckendorf and Aicha BenTaieb for fruitful discussions and feedback on this work.

\bibliographystyle{unsrtnat}
\bibliography{reference}

\begin{appendices}
\setcounter{figure}{0}    
\setcounter{table}{0}
\renewcommand{\thetable}{A\arabic{table}}
\renewcommand{\theHtable}{A\arabic{table}}
\renewcommand{\thefigure}{A\arabic{figure}}
\renewcommand{\theHfigure}{A\arabic{figure}}

\section{Related Work}
\label{sec:related_work}
\subsection{Unsupervised Image Segmentation}
Unsupervised segmentation remains an unsolved problem in computer vision, but it has shown dramatic progress in the past few years, driven largely by advances in self-supervised representation learning.
One common approach to this problem is to use saliency maps to generate object proposals or pseudo-labels to drive training.
For example, MaskContrast \cite{van2021unsupervised} uses a predicted saliency map to generate object mask proposals and then learns pixel-level embeddings by an object-centric contrastive optimization objective.
USOD \cite{zhou2023texture} addresses the dependency of MaskContrast on the quality of the saliency map, by proposing a Confidence-aware Saliency Distilling strategy to progressively distill saliency knowledge from easy samples to hard samples and refines boundaries by texture matching.

ViTs trained with SSL have also been shown to produce good representations for detecting salient foreground objects \cite{caron2021emerging, oquab2023dinov2, shamshad2023transformers}. The self-attention maps from a ViT trained with no supervision show promising results in foreground object segmentation.
TokenCut \cite{wang2022self} constructs a graph from patch-level ViT features, and then applies spectral clustering and finally graph cuts to generate a final foreground-background segmentation.
While achieving promising results, all of the above methods primarily focus on segmenting the dominant object in a scene.
However, for mucosal feature discovery from colonoscopy images, our objective is to segment tissues of different types where there is often no dominant object or clear boundaries between tissue regions, thus necessitating semantic segmentation.


Encouragingly, patch-level ViT features have been shown to localize semantic information far beyond saliency \cite{amir2021deep}.
Deep Spectral \cite{melas2022deep} is a simple yet effective method for unsupervised semantic segmentation built on this principle. 
It is closely related to TokenCut; however, rather than identifying the single most salient object in a scene, the authors apply spectral clustering to decompose the image into numerous segments, and then perform semantic segmentation by clustering these segments across the dataset.
One potential limitation of ViT-based approaches is that they learn patch-level, rather than pixel-level representations, and thus may be less appropriate for generating segmentation masks with fine detail \cite{liu2022cuts}; however, we find that this trade-off is worth it for the improved representations learned by ViTs trained with SSL, and is well suited to the task of mucosal feature discovery.

In this work, we applied the unsupervised semantic segmentation workflow from Deep Spectral due to its simplicity and efficiency. 
We also tried simpler methods, such as patch-level clustering \cite{amir2021deep} and self-attention \cite{oquab2023dinov2}, but the results were not as performant as the Deep Spectral workflow.
There are two other methods that we consider for future work.
STEGO \cite{hamilton2022unsupervised} is a related method that further enhances the clustering result by distilling self-supervised features into compact structures with a segmentation head. 
Whereas the authors in \cite{ziegler2022self} introduce a new task, self-supervised learning of object parts, and argue that this task yields better features for unsupervised segmentation.
We plan to test both methods, to see if they further improve segmentation performance.

\subsection{Self-supervised Learning in Endoscopy}
Applications of SSL in medical image analysis have led to performance boosts in a number of downstream tasks \cite{zhang2023dive, shurrab2022self}. 
For example, the Barlow Twins framework was applied to endoscopic images, leading to improved performance in classifying pathological findings \cite{nguyen2023self}. 
Endo-FM \cite{wang2023foundation} is a foundation model pretrained on a massive dataset of endoscopy videos from multiple medical sites. 
With Endo-FM, the authors demonstrated superior performance compared to other state-of-the-art methods on polyp diagnosis, detection, and segmentation.
We benchmark the performance of our internal Etro model, as well as numerous ImageNet-trained models against the Endo-FM model in this paper.
State-of-the-art SSL methods have also been systematically investigated in the context of surgical computer vision in \cite{hirsch2023self, ramesh2023dissecting}, and their results show that SSL pretraining on unlabeled videos improves the performance of surgical phase recognition (identifying the current phase of an operation) and tool presence detection.
In addition to image classification, SSL has also been used for anatomical region prediction and camera localization for endoscopy videos \cite{kelner2023semantic, yao2021motion}.

\subsection{Weakly Supervised and Unsupervised Segmentation in Colonoscopy}

Weakly supervised and unsupervised segmentation has primarily been explored for detection and segmentation of anomalies, such as polyps, in colonoscopy. 
For example, visual explanations of weakly supervised models can help localize cancerous polyps in colonoscopy images \cite{raju2023advanced}. 
In \cite{sasmal2022unsupervised}, images containing polyps were over-segmented into superpixels and then an adaptive markov random field was applied to segment polyps in an unsupervised manner. 
A domain adaptation model was proposed in \cite{xu2022deep}, that learns shared features of two domains and enables generalization of a polyp detection model to out-of-domain images. In \cite{tian2023self}, a self-supervised pre-training method was designed to learn effective
image representations for unsupervised anomaly detection.

More comprehensive mucosal feature detection and anatomical landmark detection are emerging fields.
A CNN-based method was proposed to identify anatomical landmarks from polyps and normal colon mucosa in \cite{taghiakbari2023automated}. 
In \cite{fiaidhi2022investigation}, colonoscopy frames from patients with Crohn's disease were clustered into five lesion groups using the structural similarity index.
However, none of these works attempt unsupervised semantic segmentation of colonoscopy images.
In this paper, we present the first investigation of fully unsupervised semantic segmentation for mucosal feature discovery.

\section{Datasets}
\label{sec:dataset}
\subsection{HyperKvasir Dataset}
HyperKvasir \cite{borgli2020hyperkvasir} is a comprehensive dataset of colonoscopy images collected from 2008 to 2016 in the Bærum Hospital in Vestre Viken Health Trust (Norway). It was collected using standard endoscopy equipment and contains images from both the upper (esophagus, stomach, and duodenum) and lower (terminal ileum, colon and rectum) gastrointestinal tract. 
It contains 10,662 image-level annotations, corresponding to 23 classes that are subdivided into anatomical landmarks, mucosal view quality, pathological findings and therapeutic interventions. Images with MCES labels are the most relevant to mucosal feature discovery for IBD. 
There are 201, 441, and 143 images annotated as MCES 1, 2, and 3, respectively.
There are no images annotated as MCES 0. 
In addition to image-level annotations, HyperKvasir also contains annotated polyp masks for 1,000 frames containing polyps. We adopt the published data splits for all experiments.

\subsection{Etrolizumab (Etro) Dataset}
The Etro dataset is our internal dataset containing 5,145 videos of colonoscopies and sigmoidoscopies from patients with moderate to severe ulcerative colitis in 5 clinical trials (Hibiscus I NCT02163759, Hibiscus II NCT02171429, Gardenia NCT02136069, Hickory NCT02100696, and Laurel NCT02165215) conducted across 586 different sites. It was collected using a variety of different endoscopy equipment, and the videos vary in resolution and frame rate (10 to 25 fps). We extracted frames at a rate of 1 fps and removed low-quality frames \cite{gutierrez2021training}, yielding a final dataset of 525,711 images.

\section{Implementation Details}
\label{sec:implementation}

\subsection{Training on Etro Dataset}
We trained DINO-v1 using our internal Etro dataset. All images were pre-processed by removing out-of-field regions, masking specular regions
\cite{yao2021motion}, and resizing to 256$\times$256 before being fed through the DINO pipeline. We applied the standard data augmentations proposed in DINO-v1 framework, but only augmented the image hue by a small amount to preserve color information, which is important for discriminating tissue types. In addition, we applied temporal data augmentation by sampling frames from a temporal window of approximately 0.2 seconds but did not observe a significant performance gain. The DINO model was initialized by the publicly available DINO weights pretrained on ImageNet. We explored several combinations of architectures (ViT-S, ViT-B), patch size (8, 16), base learning rates, and training epochs. The configuration with ViT-B, patch size of 16, base learning rate of 0.00025, and 100 training epochs achieved the best 23-class classification performance on HyperKvasir dataset, and this model was used for polyp segmentation and semantic segmentation. The model was trained on 8 NVIDIA A100 GPUs and training took 74 hours. All other hyper-parameters such as teacher temperature, and the number of epochs for linear learning rate warm up were default values from the official DINO-v1 repository \cite{caron2021emerging}. 

\subsection{Publicly Available SSL Models}
Publicly available SSL models were downloaded from the following Github repositories.
\begin{itemize}
    \item DINO-v1 \cite{caron2021emerging}: \href{https://github.com/facebookresearch/dino}{https://github.com/facebookresearch/dino}
    \item DINO-v2 \cite{oquab2023dinov2}: \href{https://github.com/facebookresearch/dinov2}{https://github.com/facebookresearch/dinov2}
    \item MAE \cite{MaskedAutoencoders2021}: \href{https://github.com/facebookresearch/mae}{https://github.com/facebookresearch/mae}
    \item Endo-FM \cite{wang2023foundation}: \href{https://github.com/med-air/Endo-FM}
    {https://github.com/med-air/Endo-FM}
\end{itemize}

In the main text, we refer to all models trained on natural images as ImageNet-trained models, despite the fact that DINO-v1 and MAE were trained on ImageNet-1k ($\sim$ 1.3M images), whereas DINO-v2 was trained on a new dataset ($\sim$ 142M images) containing ImageNet-22k, ImageNet-1k (train set only), Google Landmarks, and curated images from a publicly available repository of crawled web data.

\subsection{Image Classification}
To evaluate the quality of image-level features from trained SSL models, 
we calculated the performance for linear probing and weighted nearest neighbor (KNN) classifiers.
For both classifiers, we performed a two-fold cross-validation using the official HyperKvasir split and reported the averaged micro F1 and macro F1 score. 

For linear probing, a single linear layer was added to each model and trained on the labeled images. We tuned the following hyperparameters: (i) average pooling of the patch token (ii) the number of ViT blocks to extract embeddings from, and (iii) the learning rate. We used Stochastic Gradient Descent (SGD) optimization to minimize the cross-entropy loss for 100 training epochs. 

For the KNN classifer, the feature of an image in the validation set was matched to the k nearest features in the training set. We experimented with k = 15, 20, 25, 50 and found that k = 20 tended to have good performance across different models and tasks (similar findings were reported in \cite{caron2021emerging}).

\subsection{Semantic Segmentation using Deep Spectral}
To extract segmentation masks, we first passed input images through our pretrained ViTs and extracted patch-level features from the last four ViT blocks.
We then calculated the feature affinity matrix from the patch-level feature correlation, and the color affinity matrix using a down-sampled image and the sparse KNN-matting matrix \cite{melas2022deep}.
The final affinity matrix is a sum of the feature affinity matrix and the color affinity matrix. We performed spectral clustering on the affinity matrix to decompose each single image into semantically meaningful segments. The number of segments was adaptively determined based on the magnitude of eigenvalues. 

Next, we cropped the region around each segment and fed them back into the same ViT. The patch embeddings from the last four blocks were extracted and averaged to represent the whole segment. After that, we applied principal component analysis (PCA) to reduce the feature dimension and then performed K-means clustering across a collection of images. Semantic segmentation masks were built by assigning each segment with the corresponding cluster ID from K-means.

We found that the results of unsupervised segmentation were sensitive to the choice of hyperparameters. 
Three types of hyperparameters were explored (i) image preprocessing, including image normalization using adaptive histogram equalization, and image resizing; (ii) the weight parameter for the color affinity matrix, which balances the semantic and low-level color consistency; and (iii) the choice of K in K-means. To identify a reasonable combination of parameters, we randomly selected 10 images and qualitatively assessed the results from spectral clustering and K-means clustering. From our experiments, we found that the color affinity matrix helped to break a single image into more semantically meaningful segments. While the performance of polyp segmentation didn't benefit from the color affinity matrix, it did aid in generating more interpretable concepts in mucosal feature discovery, with a weight of 1.0.

The performance of polyp detection and segmentation was evaluated using two-fold cross-validation. For unsupervised polyp segmentation, we applied K-means to the training set, and then visualized the predicted semantic segmentation mask from 10 randomly selected frames in the training set to identify the cluster corresponding to polyps. Subsequently, we applied the learned PCA model and K-means model directly to the validation set and compared the resulting segmentation mask from the identified cluster with the ground truth. 
For linear probing, we trained a linear classifier on segments from the training set, where segments were labeled by whether or not they are from a polyp. Then the linear classifier was applied to the validation set and generated binary polyp segmentation masks based on the classification results.

For polyp detection, we reported the F1 score at IoU=0.3. 
This metric defines a true positive as a predicted mask with an IoU $\geq$ 0.3 compared to the ground truth mask, a false positive as a predicted mask with an IoU $<$ 0.3 compared to the ground truth mask, and a false negative as an annotated polyp that the model failed to detect.  For polyp segmentation, we reported the average IoU values.

\section{Additional Results}
\label{sec:add_results}

\begin{table}[]
\caption{\textbf{Performance of KNN Classifiers on Image Classification.} We report the average macro-F1 and micro-F1 scores from a two-fold cross-validation.}
\label{tab:KNN}
\begin{tabular}{llcccc}
\hline
\multirow{2}{*}{Type of method}                                                            & \multirow{2}{*}{Model name} & \multicolumn{2}{c}{Full 23-class} & \multicolumn{2}{c}{MCES 3-class} \\
                                                                                           &                             & F1 (ma.)        & F1 (mi.)        & F1 (ma.)        & F1 (mi.)       \\ \hline
\multirow{2}{*}{\begin{tabular}[c]{@{}l@{}}Domain SSL \\ w/ KNN classifier\end{tabular}}   & DINOv1 VitB/16 (Etro)      & 52.1            & 81.2   & 61.4            & \textbf{67.5}  \\
                                                                                           & DINOv1 VitB/16 (Endo-FM)   & \textbf{53.2}   & 83.0            & \textbf{62.5}   & 66.7           \\ \hline
\multirow{4}{*}{\begin{tabular}[c]{@{}l@{}}ImageNet SSL \\ w/ KNN classifier\end{tabular}} 
                                                                                           & DINOv1 VitB/8              & 52.7            & \textbf{84.5}   & 58.7            & 66.7           \\
                                                                                           & DINOv1 VitB/16             & 51.9            & 83.7            & 56.2            & 64.4           \\
                                                                                           & DINOv2 VitL/14             & 48.6            & 79.9            & 49.1            & 62.0           \\
                                                                                           & MAE VitL/16                & 47.3            & 79.3            & 51.7            & 61.9           \\ \hline
\end{tabular}
\end{table}

\begin{figure}
\centering
\includegraphics[width=13cm]{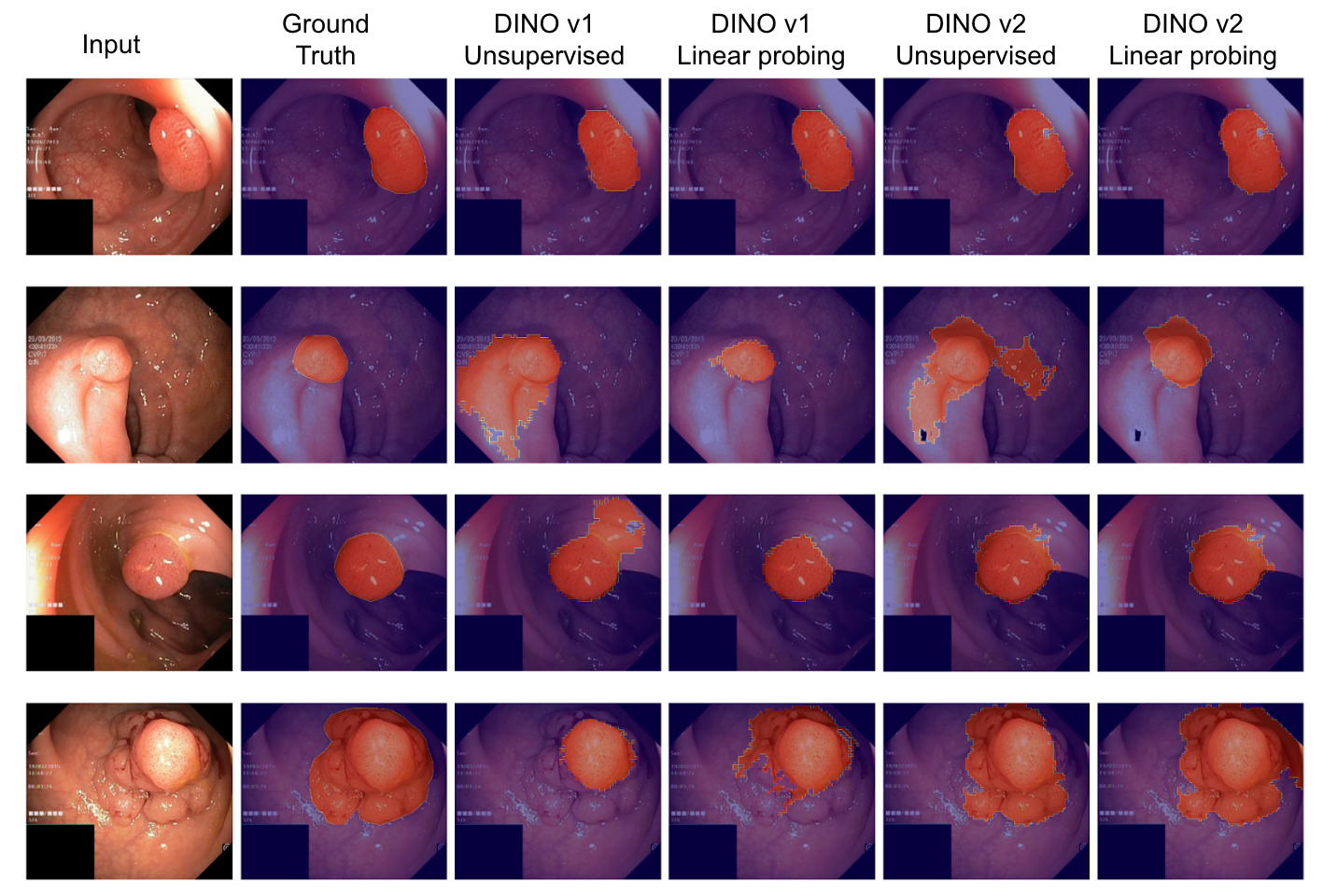}
\caption{\textbf{Polyp Segmentation Examples.} The first column shows input images, the second column shows ground truth masks overlaid on input images, and the remaining columns present predicted polyp masks from different methods overlaid on input images. The predicted masks in this figure are generated with ImageNet-trained models (DINOv1 VitB/8, DINOv2 VitL/14). }
\label{fig:polyps}
\end{figure}



\begin{figure}
\centering
\includegraphics[width=13cm]{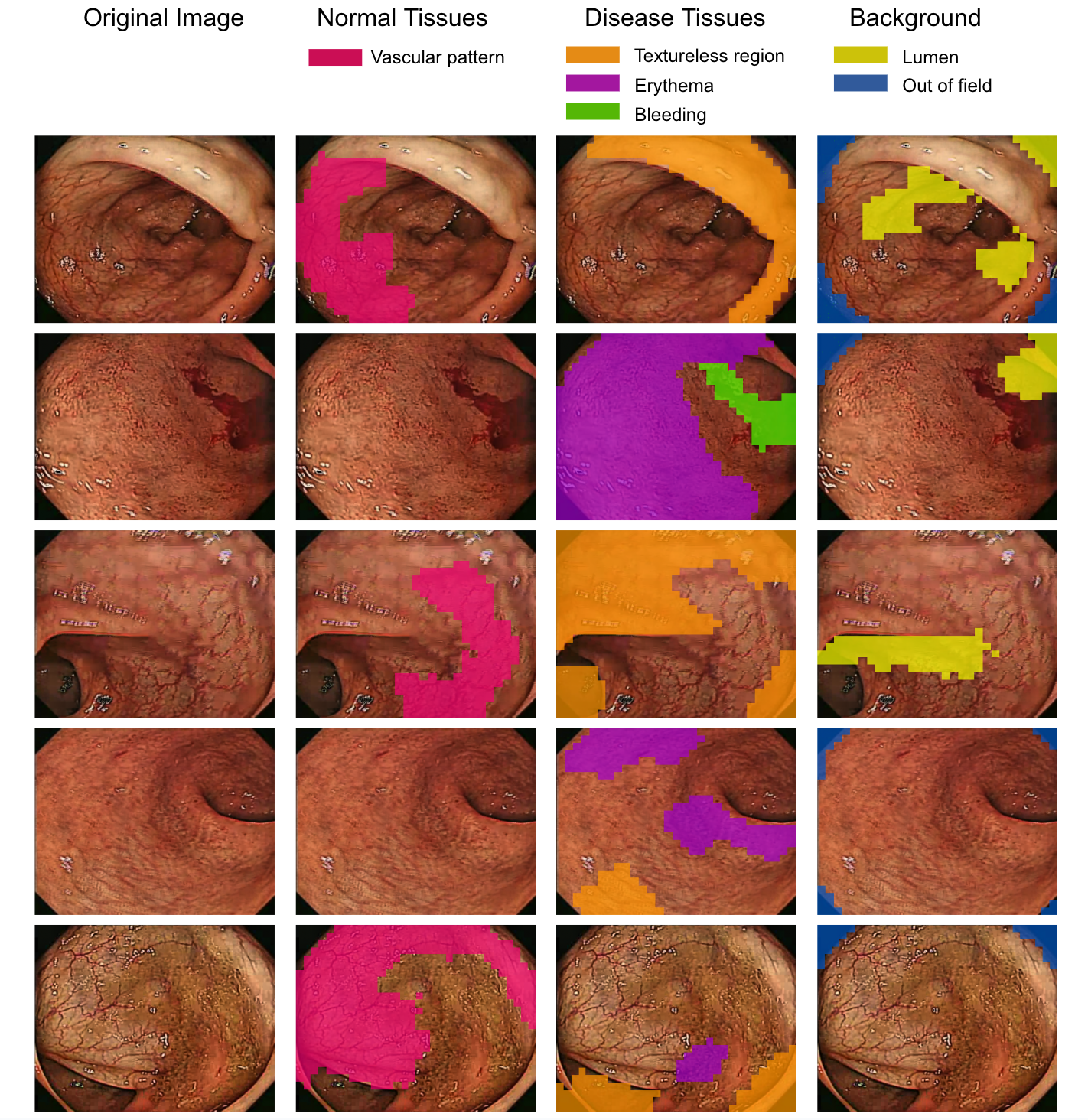}
\caption{\textbf{Unsupervised Discovery of Mucosal Features.} We show additional examples of images overlaid with `discovered' concepts related to normal tissues, disease tissues, and background, respectively.}
\label{fig:semantic_seg_more}
\end{figure}

We present the following additional results:
\begin{enumerate}
    \item Performance of KNN classifiers on image classification tasks are presented in Table \ref{tab:KNN}
    \item Polyp segmentation results are presented in Figure~\ref{fig:polyps}. 
    \item Additional examples of mucosal feature discovery are presented in Figure \ref{fig:semantic_seg_more}.
\end{enumerate}



\section{Limitations and Future Work}
\label{sec:limitation}
The focus of this study was to evaluate the feasibility of unsupervised segmentation for mucosal feature discovery, and the role of SSL representations in this task.
As such, we adopted the Deep Spectral workflow as a simple and strong baseline for unsupervised segmentation from ViT patch-level features.
However, there are numerous opportunities for improvement.
(1) We used patch-level features from the last four layers, following the protocol in DINOv2; however, existing research suggests that utilizing key / values \cite{amir2021deep,melas2022deep} from attention layers may lead to better segmentation performance. Additionally, the spatially dense clustering task proposed in \cite{ziegler2022self} could potentially enhance the quality of SSL representations for image segmentation. 
(2) Spectral clustering could be replaced by more sophisticated methods such as diffusion condensation \cite{brugnone2019coarse, liu2022cuts}.
(3) For the final segmentation, K-means was chosen due to its simplicity and efficiency. 
In the context of feature discovery, methods such as hierarchical clustering may prove advantageous. 

We also observed that the results from unsupervised segmentation were sensitive to hyperparameter choices, such as the weights of color affinity and feature affinity in spectral clustering, the number of clusters for K-means, etc. 
Future work could investigate a systematic method for parameter tuning with appropriate metrics.
For example, a few salient and well-known mucosal features (e.g. bleeding) could be annotated in a small number of samples and serve as a guide for tuning the segmentation parameters.

During this study, we evaluated the performance of both ImageNet SSL models and domain-specific SSL models in three downstream tasks. However, no single model consistently outperformed the others. In general, we observed that ImageNet SSL models seemed to excel at distinguishing structure-related tasks, while domain-specific SSL models performed better at identifying subtle differences in mucosa. 
Future work could explore a broader range of architectures, training strategies and downstream tasks to gain a deeper understanding of this relationship. 
Furthermore, a double-blinded review by domain experts could provide a more unbiased evaluation of the quality of segmentation masks generated by the different methods explored in this work.
Finally, and perhaps most importantly, we plan to investigate the correlation of mucosal features from unsupervised semantic segmentation with patient-level clinical outcome metrics.


\end{appendices}

\end{document}